# CReIS: Computation Reuse through Image Similarity in ICN-Based Edge Computing

Atiyeh Javaheri, Ali Bohlooli*, Kamal Jamshidi

*Faculty of Computer Engineering, University of Isfahan, Isfahan, Iran*
*Corresponding author:Ali Bohlooli*
*E-mail address: a.javaheri@eng.ui.ac.ir (A. Javaheri), bohlooli@eng.ui.ac.ir (A. Bohlooli), jamshidi@eng.ui.ac.ir (K. Jamshidi).*

**Abstract**

At the edge, there is a high level of similarity in computing. One approach that has been proposed to enhance the efficiency of edge computing is computation reuse, which eliminates redundant computations. Edge computing is integrated with the ICN architecture, capitalizing on its inherent intelligence to facilitate computation reuse and reduce redundancies in computing operations. In many past works, ICN's ability to enable computation reuse through caching has been limited. In this context, a new approach is proposed that considers computation requests with similar input data, which yield identical results, as equivalent. This method facilitates computation reuse through caching in ICN. The use of approximate results to reduce redundant computations without requiring high accuracy in input matching is provided. This concept is termed the "Similarity Index", which effectively considers images to be similar despite minor changes in the angle of photography. The Similarity Index is determined through an algorithm known as HNSW and utilizes the SIFT descriptor to identify similar data. This approach helps reduce user latency times by providing quick access to results. The evaluation, simulated using the ndnSIM tool, showed an 86% improvement in completion time compared to scenarios without computation reuse, whereas previous works reported only a 70% improvement. To strengthen this method, an analytical model for computing request transfer considering computation reuse in ICN-based edge computing is provided. To assess the accuracy of the model, several evaluations have been conducted in the simulator by varying the parameters, resulting in a maximum error percentage of approximately 16%.

## 1. Introduction

Cloudlet, introduced by Satyanarayanan, is a prominent classification within the realm of edge computing. Its primary objective is to mitigate latency by strategically positioning computing resources in close proximity to end users [1, 2]. A cloudlet can be described as a cluster of servers between the cloud and mobile devices. They act as independent cloud environments at the edge and are able to provide services even when the connection to the central cloud is disrupted. However, compared to cloud data centers, they have low scalability and limited resources [3]. Cloudlets are typically deployed in public places such as coffee shops, libraries, and smart cities where multiple users and devices are in close proximity. These environments demand high scalability to handle the substantial computing task generated by the large number of users and devices. Therefore, low-scalability edge computing must support high-scalability use cases. Additionally, in these scenarios, the close proximity of devices results in a high number of duplicated tasks, as many services are similar and produce the same computational results [4, 5]. The execution of each of these duplicated tasks leads to a large amount of redundant (extra) computations, which raises concerns about redundant computations at the edge. Therefore, the approach of " Computation Reuse " is employed. Computation reuse is a term that involves the storage and utilization of previously computed results, thereby avoiding the computation of redundant tasks. Computation reuse can be partial or full. Previous computing results are readily available, thus reducing completion time. Additionally, it is possible to prevent computational overhead for the same tasks and decrease resource capacity limitations. On the other hand, many similar services differ only in the angle of the image or the lighting in the input data, which produces the same computation results. There is no proper comparison to distinguish them, and they are considered different. It is important that these data are considered the same, so we define a new factor called the "Similarity Index". This factor is meant to detect more similarities for input data images that have different angles but will yield the same results. The Similarity Index can be very important, especially in cases where computations are very time-consuming, as many computations are similar, resulting in reduced completion time.

Computation reuse in edge computing employing the TCP/IP architecture encounters various challenges. One of these challenges is the computation of redundant tasks. Because routers do not provide caching and are not aware of the requests and responses, therefore routers cannot identify the same or similar tasks and forward the same requests to the edge server for computation reuse. To solve the challenge of enhancing computation reuse, it is proposed to integrate edge computing with information-centric networks (ICN). ICN places infrastructure at the edge of computing and enables similar requests to be forwarded to the same edge server for reuse, either fully or partially [6]. ICN can identify similar tasks by naming and store the computation results by caching. However, traditional ICN cannot be directly deployed at the edge to leverage its caching capabilities. This presents several challenges, including computation reuse, resource discovery, security management, mobility, and service invocation [4,7-11].

In prior research, the emphasis on computation reuse in ICN has predominantly focused on the forwarding features and naming conventions. These strategies send similar requests along the same path, leading to improved computation reuse [4,7- 10]. All of these papers ignore the potential of the cache feature, but the last two focus specifically on caching. In our previous paper [12], we used a CS table change, while in [4], locality-sensitive hashing (LSH) was employed to utilize the ICN cache. In these studies, the number of computation reuses is low compared to the overhead. Our approach to computation reuse involves utilizing the Similarity Index through an algorithm called the Hierarchical Navigable Small World (HNSW). Recognizing the importance of computing reuse, this paper introduces a new analytical model to represent the computing transfer process in Named Data Networking (NDN)-based Edge Computing.

The main contributions of our article are as follows:

•We have introduced a novel factor called the Similarity Index, which enables the computation reuse of input data images with different angles. The goal is to be able to identify similar tasks whose computational results are the same. Consequently, this factor facilitates quick access to results and reduces delays for users.

•The Similarity Index utilizes the HNSW algorithm, which is employed by the Scale-Invariant Feature Transform (SIFT) descriptor algorithm to identify similar tasks.

•Due to the identification of similar tasks, it utilizes ICN cache, enabling users to reach the computation results sooner. As a result, the completion time is reduced.

•We have developed a new analytical model that accurately models the computation transfer process in NDN-based edge computing, with a specific emphasis on computation reuse.

•Through extensive simulations conducted under diverse settings, our proposed scheme has yielded valuable insights and results, validating its efficacy and practicality.

The paper is organized as follows:

Motivation and background are explored in section 2, while section 3 focuses on related works. The proposed method is presented in section 4, followed by the evaluation in section 5. Finally, section 6 provides a summary and discusses future works.

## 2. Motivation and Background

### 2.1 Motivation

In cloudlet scenarios, the proximity of devices leads to an increase in repetitive tasks, as many services are similar and produce the same computational results. Executing each of these repetitive tasks results in performing a significant amount of redundant computations, raising concerns about additional computations due to reduced scalability at the edge. Therefore, leveraging the concept of approximate similarity in computation requests allows the system to reuse cached results instead of executing each task from scratch. This approach enhances time efficiency and reduces energy consumption. Therefore, the main goal of this paper is to present an approach for computation reuse in the cache in ICN-based edge computing by considering image similarity. Computation reuse across images that are time-consuming or complex can be highly beneficial. Furthermore, in real-world scenarios, identical requests for images are typically rare or even non-existent. To facilitate this process, the HNSW algorithm is employed, enabling the identification of similar tasks and enabling caching mechanisms. This approach not only significantly reduces completion time but also minimizes computational costs.

Additionally, a novel approach for modeling the computation transfer process in NDN-based edge computing is introduced. Simulations have shown that the proposed model has a low error rate.

## 2.2 Background

### 2.2.1 NDN Architecture

The underlying framework of the NDN architecture revolves around a pull model. This model necessitates the transfer of two distinct packets, namely the request (interest packet) and the data packet containing the requisite information, in order to facilitate communication between the producer and the user. The NDN is one of the prominent architectures within ICN.

Each router within the NDN framework upholds three essential data structures, namely the Content Store (CS), Pending Interest Table (PIT), and Forwarding Information Base (FIB). These structures help manage and route the flow of interest and data packets within the network [13].

Upon reception of a packet of interest, each router initially examines its local CS table to discover a complete correspondence of all the prefixes of interest. The search criteria rely on a nomenclature method where only precise matches are considered valid. In the event of a perfect match, the data packet is transmitted to the supplicant through the same interface as the interest packet was received, and subsequently, the interest packet is discarded. Conversely, if no precise match is found, the PIT entry is assessed. If a match is discovered in the PIT entry, it signifies that the interest packet has already been received and sent upstream. Subsequently, the incoming interface is appended to the corresponding PIT entry, and the new interest packet is disregarded. Conversely, if the request is novel, a fresh entry is generated in the PIT, which encompasses the name of the interest and the incoming interface. Ultimately, the FIB entry is examined to ascertain the longest prefix match and to transmit the subsequent hop towards the producer. In the event that no match is identified in the FIB entry, the interest packet is discarded. When the content is found, the router sends the data packet to the requester [13].

When a router receives a data packet in the reverse path, the first step it takes is to examine its PIT entry. In the event of a match, the data packet will be forwarded to all interfaces that have been saved in the PIT table. Conversely, if no match is found, the data packet is discarded. As the data packet progresses along the data path, it has the possibility of being stored in the CS table based on the cache policy. Storing the data in the CS table enables subsequent requests to be satisfied from the cache, thereby reducing the necessity of retrieving the data from its original source. This caching mechanism plays a significant role in enhancing the efficiency and performance of content retrieval and delivery within the NDN architecture [13].

### 2.2.2 HNSW Algorithm

HNSW is an approximate nearest neighbor search algorithm designed to efficiently find similar items in high-dimensional spaces. This algorithm is an evolved version of the Navigable Small World (NSW) algorithm and Probability skip list, utilizing an adjacency list. It features a hierarchical structure with multiple levels, each containing a different number of nodes, where each node represents an item or data point. The upper layers contain fewer points and longer edges for faster searching, while the lower layers contain most points and shorter edges for more precise searches [14].

To insert a data point, the maximum number of layers allowed for insertion is determined by an exponentially decaying probability distribution.

The HNSW algorithm has two parameters named M(Mmax) and EF (efConstruction). The variable M(Mmax) controls the maximum number of neighbors that can be connected in each layer of the graph structure. The insertion process ends when the connections of the inserted elements reach the specified M(Mmax) value. The efConstruction parameter (which represents the graph size) controls the structure and determines the search size during the nearest neighbor search process. A dynamic list W of efConstruction is maintained throughout the search [14].

In the first stage of the search, the algorithm greedily traverses the graph from the top layer to select the nearest neighbors. It then continues the search from the next layer, using the nearest neighbors found in the previous layer as input points, and this process is repeated [14].

### 2.2.3 Use-case

This method is suitable for use cases in which users have access to images, and edge computing performs computations on those images. For example, two cases are mentioned:

**Smart Tourism with Augmented Reality (AR)**: In this scenario, as depicted in Fig. 1, visitors take photos of a target object from different angles, lighting, and distances using their mobile devices to obtain more information and annotations. They then send these images to the edge server for computing. Although the submitted images may be different, the end result of the information presented to the visitors will be the same.

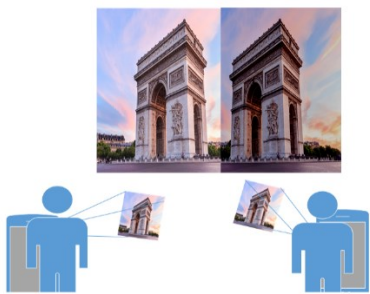

**Fig. 1.** An example of smart tourism with AR

**Autonomous Car:** Consider a scenario where autonomous car sensors capture images to gather information. In such a scenario, multiple autonomous cars may request the same computational tasks at the same time or place, which will produce the same results. For example, in a specific location, such as with traffic signs, images are taken with different angles and lighting. Between two consecutive snapshots at the same time, there may be an overlap, and computing the same results is unnecessary. Considering the amount of redundant computation, computation reuse can effectively improve the performance of autonomous cars.

## 3. Related Work

Recent studies have indicated that ICN-based edge computing has the potential to offer a highly efficient network. Consequently, multiple frameworks have been devised to seamlessly integrate ICN with edge computing. The pioneering effort to enable computational capabilities through NDN is referred to as Named Function Networking (NFN) [15].

The use of ICN in edge computing and its transition from an information-centric framework to a computation-centric framework pose several challenges. These challenges include resource discovery, mobility management, security management, and computation reuse. For example, in [6], a proposition is made for an ICN-based edge computing service model that incorporates a naming scheme, service session model, and forwarding strategies. Meanwhile, [11] introduces an approach for service discovery and invocation as well as user mobility management.

This paper exclusively focuses on methods for computation reuse. Therefore, we will continue to review past work in the field of computation reuse.

ICedge leverages computation reuses through naming conventions and forwarding schemes. Each naming convention is linked to a forwarding scheme that enables the reuse of previous computations by directing similar requests to the same Compute Node (CN). The forwarding scheme includes CONFIG, CATEGORY, and ZONE. The ZONE forwarding scheme groups tasks based on regions, locations, or sections of a map, following the naming convention of "X_coordinate/Y_coordinate/". The CATEGORY forwarding scheme groups tasks based on their building and floor locations, using the naming convention of "/building_X/floor_y/Room_z" and the naming convention "/config_X/Param_Y" encompasses a wide array of parameters for the matrix multiplication service [7].

A proposed algorithm for computation reuse has been introduced that focuses on a forwarding strategy [10]. In addition to the PIT and FIB tables, there is an additional table called the Forwarding Task Table (FTB). The FTB table includes the task name and the interface used for forwarding. To achieve computation reuse, the router searches the FTB table, resulting in the computation of a similarity score. The determination of this score is based on the longest prefix match of the task name.

In [9], three forwarding strategies—passive statistical, active probing, and active-passive hybrid—have been proposed. These forwarding strategies aim to eliminate the need for additional computation, reduce service completion time, and achieve load balance. In the passive statistical forwarding strategy, the cost of the service is recorded for each interface, calculated as the ratio of the service completion time to the required resources. The service completion time encompasses both the time needed for network transmission and computing. When forwarding, the interface with the most cost-effective service is selected. In the active probing forwarding strategy, the acquisition of data regarding the status of network links, available resources, and the load information of edge servers is accomplished through the transmission of probe request packets. The estimation of service completion time is made possible through the utilization of this aforementioned data, as well as the round-trip time of a probe. Consequently, the interface with the minimal anticipated service completion time is selected. This strategy incurs substantial probing overhead costs. Therefore, an active-passive hybrid forwarding strategy has been proposed. This strategy achieves the extraction of information pertaining to the resources and load information of the edge server through a confirmation packet.

Previous works have proposed mechanisms that incorporate naming and forwarding techniques to direct service requests toward suitable CNs. However, none of them emphasized the concept of computation reuse through caching. This paper continues to review computation reuse through caching.

In [4] a framework called Reservoir is presented with the aim of achieving pervasive computing in the edge network. This naming and routing framework extends NDN by incorporating LSH and builds a compute-aware reuse network, so that similar tasks are identified in a lightweight manner in the network and similar tasks are routed to the same Edge Nodes (ENs). The presence of LSH in the Reservoir namespace has led to the utilization of caching and request aggregation features in NDN, enabling computation reuse in the cache.

In terms of computation reuse through forwarding, a new data structure called rFIB has been created by extending the NDN routing pipeline. After receiving the request, the NDN router performs a search in the cache. If a miss occurs, the request is inserted into the PIT table, and the router then looks up the rFIB, selecting one of the bucket entries based on the highest similarity among previously executed tasks, facilitating computation reuse. The LSH algorithm is based on similarity storage; however, its overhead and search speed are high.

In our previous work [12], computation reuse is achieved through the CS table modification method. The CS table is altered so that identical input data with different names are considered the same, allowing us to utilize the results of previous computations. Additionally, this approach is suitable for non-image input data of small size. However, if there are many input parameters, it can incur additional overhead for storage.

## 4. Proposed Method

In this paper, the identification of similar input data is achieved by employing the HNSW algorithm. The primary reasons for selecting the HNSW algorithm for ICN-based edge computing are its scalability and efficient search capabilities, which result in low latency.

The input data requests for services in this study consist of images, where the image names are appended to the end of the naming convention. For example, the format is "/<service-name>/<ID (unique)>/<image-name>", with image names sourced from the MNIST dataset. By adhering to this naming convention, computation requests are formulated and transmitted across the network to initiate the desired computations.

The workflow of the interest operation is illustrated in Fig. 2. First, the image names requested (the third part of the naming convention) are extracted and loaded into the OpenCV library. Each image is read in grayscale using cv::imread with the cv::IMREAD_GRAYSCALE flag and resized to 50 by 50 pixels, ensuring that all images have a fixed size for descriptor extraction. Image descriptors are then obtained using the SIFT algorithm.

Descriptors are extracted from the images using key points, which represent distinctive and prominent locations within the image and capture crucial visual information. Each image's key points correspond to a specific descriptor.

The SIFT descriptor offers accurate and robust identification against variations in lighting and background, as well as resistance to rotation and minor changes in image angles. These features are often useful in image recognition and image similarity applications.

For each image, descriptors are computed using the SIFT algorithm through sift->detectAndCompute. The resulting descriptors are then reduced to one dimension using the cv::reduce function with the cv::REDUCE_MAX flag, which selects the maximum value from each row. This step effectively condenses multi-dimensional data into a one-dimensional format.

Descriptors were then converted into float format and provided as input to the HNSW algorithm. This algorithm is implemented using the hnswlib::HierarchicalNSW class, which offers the capability to build and manage a hierarchical graph structure for approximate nearest neighbor searching. Each descriptor stored in cache is added using the appr->addPoint function. The input descriptors serve as query points to search for the nearest neighbors in the HNSW. This is accomplished using the alg_Hnsw->searchKnn function, which performs a k-nearest neighbor search based on the similarity of the descriptors. The parameter k determines the number of nearest neighbors to retrieve. By comparing the image descriptors, the similarity or dissimilarity between images can be assessed.

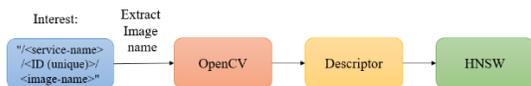

**Fig. 2.** Workflow of the interest operation

The first request is sent to the CN and cached descriptor image and computing result in the reverse path. For subsequent requests, the image name descriptors are compared using the HNSW algorithm to evaluate the similarity between the requested data and the data in the cache. They compute the degree of similarity between the data and find the best one for the input request data. In other words, by evaluating the similarity of the descriptors to each other, this method looks for the descriptor similar to the request that has already been computed and placed in the cache. The similarity of descriptors allows input data from images with different angles, which appear the same to the human eye, to be treated as the same, even if their naming conventions differ. The descriptor of each input image, serving as query points, is compared to search for the nearest neighbor among the cached image descriptors. If a good match is found in the cache, the corresponding data result is returned as the query result. Otherwise, if no match is found, the request is forwarded to the next router to reach the edge. At the edge, computations may be implemented from scratch or through partial or full reuse of computation.

The design of our method is illustrated in Fig. 3, where HNSW is placed in the cache. Next, to strengthen the work, we establish an analytical model for the process of sending a computation request and transferring computation results within the ICN-based edge network. This model takes into account the potential for computation reuse and involves determining the costs associated with transmission and computations, along with performance metrics such as virtual round-trip time (VRTT) and cache miss rate at node i within a cascaded topology.

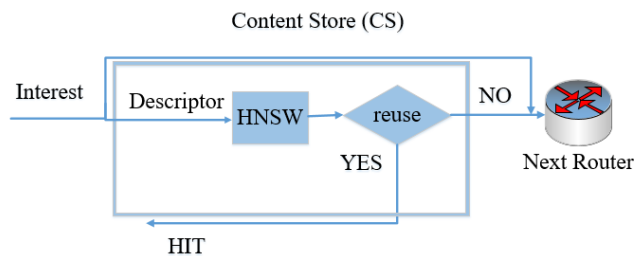

**Fig. 3.** Computation reuse design using HNSW

### 4.1 System Model and Equation Presentation

In this section, we introduce the system model and outline the problem equation. The goal is to present an analytical model of the process from sending a computation request to transferring computation results in an NDN-based edge network. For ease of reference, the symbols representing the system parameters are summarized in Table 1.

#### 4.1.1 System Model

We are considering a system with an independent edge network ($E$), which includes end users (denoted as $g$ for USER), a set of ICN routers (represented as $q$ for CS router), and compute nodes (abbreviated as $n$ for CN). The compute nodes offer users a range of pre-installed services without requiring prior configuration or registration. Within the network ($E$), there are $Y$ services, denoted as $S = \{s_1, s_2,..., s_Y\}$. Each CN contains a set of these services, such that $S = U_{i=1}^{n} s_i$.

Each service is decomposed into a set of atomic services (denoted by $K$), such that $s = \{a_1, a_2,..., a_k\}$ where $s \in S$. Users can connect to the edge network using mobile devices, and the CN offer M types of computations, which can be categorized into $K$ classes, with each class containing $m = K/M$ types of computations. The probability of different computation requests for each class is determined using the Zipf probability distribution function. The Zipf Distribution is capable of determining the popularity of content on a digital media platform.

The general form of the Zipf probability distribution function for requesting content in the entire network is given by Equation (1), where $k$ represents the class parameter, $M$ is the computation diversity, and α denotes the popularity distribution. Increasing the value of α results in a higher probability of popular computations being requested. The contents within the class are requested with the same probability $q_k$ according to the Zipf distribution, which can be simplified as $q_k = c/k^\alpha$ for convenience, where $c$ is a positive value [21].

$$P(k) = \frac{\frac{1}{k^\alpha}}{\sum_{i=1}^{M} \frac{1}{i^\alpha}} \tag{1}$$

The Least Recently Used (LRU) cache replacement policy is employed due to its low complexity, and its caching operation has a linear speed [22]. The popularity distribution in the $i - th$ node can be expressed as Equation (2) in the case of a cascaded topology [22].

$$q_k(i) = \frac{\prod_{j=1}^{i-1} p_k(j) q_k}{\sum_{l=1}^{k} q_l \prod_{j=1}^{l-1} p_l(j)} \tag{2}$$

**4.1.2 Task Model**

Each computation request (task) involves invoking a service and includes the following components: the size of the input data ($input_{tsk}$ in bits), the size of the output data ($output_{tsk}$), the workload ($w_{tsk}$), and the task's execution time ($\Delta_{tsk}$). The workload is expressed as the amount of floating-point operations per second (FLOPs), representing the required computational cost (i.e., the number of floating-point operations) that the system must perform to execute the task.

Each task can be divided into a set of sub-tasks (up to $H$) $tsk = \{sub\_tsk_1, sub\_tsk_1, ... sub\_tsk_H\}$ that have certain dependencies on each other. Each subtask calls the atomic service and includes parameters such as $\Delta_{sub\_tsk}$, $w_{sub\_tsk}$, $output_{sub\_tsk}$, and $input_{sub\_tsk}$, such that $w_{tsk} = \sum_{i=1}^{H} w_{sub\_tsk(i)}$.

Dependency information, also known as a task Directed Acyclic Graph (DAG), can be used for partial reuse of computations. Subtask dependencies are represented by a DAG $Gt(vt, \xi)$, where $vt$ is a set of subtasks and $\xi$ is a set of links between adjacent nodes, with $\xi = \{(i, j)|i, j \in tsk, i \to j\}$ describing the dependency between subtasks.

This means that subtask $j$ can only be executed after subtask $i$ is completed, indicating that the output of $i$ is required as input for $j$.

In fact, subtask dependencies should be divided in a way that they become independent so that they can be transferred to CNs in parallel. If subtasks are completely independent of each other and do not need each other's output, then using a DAG for managing dependencies is not necessary. Otherwise, using a DAG ensures that subtasks are executed in the correct order. This approach helps identify common computations among subtasks and enables their reuse.

**4.1.3 Communication Model**

This model shows how the cost of communication between computation requests and the cache is determined based on bandwidth and the number of computation requests. A task can be executed at the edge or answered by the cache, which may be fully or partially reused. The communication cost when executing a computation request is equal to the ratio of the input and output data transfer (measured in bits) to the bandwidth. The communication cost required for executing a task in CN and with full cache reuse and partial cache reuse are illustrated in Equations (3), (4), and (6). If a computation request is already stored in the cache and is fully reused, the communication cost is equal to the ratio of the input and output data transfer (in bits) to the cache bandwidth. In the case of partial cache reuse, it is assumed that $f$ sub-tasks are executed in the cache, while the remaining sub-tasks (denoted as $f'$) are executed in CN, and the sum of the subtasks is equal to $H$. Consequently, Equation (5) is derived. In this Equation, the total communication cost for $f$ sub-requests in the cache and $f'$ in the CN is computed. This cost is equal to the sum of the ratio of input and output data transfer ($f$) to the cache bandwidth and the ratio of input and output data transfer ($f'$) to the bandwidth of CN. $b_{CS}$ and $b_{CN}$ represent the bandwidths of CS and CN, respectively.

**Table 1:** Summary of modeling symbols

| Title | Description | Title | Description |
|---|---|---|---|
| $s$ | Atomic services | $S$ | Services |
| $M$ | Total number of computation types | $w_{tsk}$ | Workload task |
| $K$ | The number of computation classes | $C_{CN}$ | Computing system |
| $m$ | The number of computations per class | $\beta'$ | The cost of remaining task execution |
| $P_S$ | The possibility of full reuse of computations (Similarity Index) | $\beta$ | The cost of running the task from the beginning |
| $\alpha$ | Zipf distribution parameter | $P_{CN}$ | The possibility of executing the task in CN |
| $\sigma$ | Average number of packages | $l$ | Search cost |
| $P$ | Segment length in bytes | $\rho$ | Usage rate in CN |
| $Q$ | Generator | $p_n$ | The Equilibrium Probabilities |
| $q_k$ | Popularity of content belonging to class k | $\lambda$ | The overall entry rate of requests |
| $q_k(i)$ | Popularity of content belonging to class k in node i | $\mu$ | Service rate |
| $H$ | The sum of the subtasks | $w$ | Waiting time |
| $\alpha_\tau^{CN}$ | CN communication cost | $X$ | Router cache size |
| $R_\tau^{CN}$ | Computation reuse cost in CN | $R_\tau^{CS}$ | Computation reuse cost in CS |
| $\alpha_\tau^{CS}$ | Cs communication cost | $P_k(1)$ | Cache miss rate in the first level |
| $\alpha_\tau^{part'}$ | The communication cost of partial reuse of computations in cache | $P_k(i)$ | Cache miss rate of class $k$ in node $i$ |

$\tau$ denotes the time interval.

$$\alpha_\tau^{CN} = \frac{input_{tsk} + output_{tsk}}{b_{CN}} = \frac{\sum_{i=1}^{H} input_{sub\_tsk(i)} + \sum_{i=1}^{H} output_{sub\_tsk(i)}}{b_{CN}} \quad (3)$$

$$\alpha_\tau^{CS} = \frac{input_{tsk} + output_{tsk}}{b_{CS}} = \frac{\sum_{i=1}^{H} iutput_{sub\_tsk(i)} + \sum_{i=1}^{H} output_{sub\_tsk(i)}}{b_{CS}} \quad (4)$$

$$f + f' = H \quad (5)$$

$$\alpha_\tau^{part\prime} = \frac{\sum_{i=1}^{f} input_{\text{sub\_tsk(i)}} + \sum_{i=1}^{f} output_{\text{sub\_tsk(i)}}}{b_{CS}} + \frac{\sum_{i=1}^{f\prime} input_{\text{sub\_tsk(i)}} + \sum_{i=1}^{f\prime} output_{\text{sub\_tsk(i)}}}{b_{CN}} \quad (6)$$

### 4.1.4 Computation Model

The computation model for task execution time involves several scenarios:
1) Complete execution at the edge, where the task is entirely executed by CN (Equation 9).
2) Full or partial reuse of computations in CN (Equation 10).
3) Full or partial reuse of computations in the cache (Equation 11). The computation cost is illustrated for each case. The cost of running the task from the beginning is denoted by β, which is obtained by dividing the complexity of the task by the computing capacity. The computational capacity in CN is represented as $C_{CN}$. The complexity of the work is obtained by dividing the workload by the total work deadline from Equation (7).

The parameter of the Similarity Index indicates the probability of full reuse. According to the Similarity Index, the higher this factor is, the higher the probability of full reuse. The probability of full reuse is denoted as $P_S$, and the cost of execution, which is only for searching, is denoted as $l$.

The probability of partial reuse is $(1 - P_S)$. If the reuse of computations is partial, in addition to searching, the cost of continuing the execution for the subtasks in CN is also considered. This means that the number of $f\prime$ sub-tasks must be executed in CN. The cost of the remaining execution of the task is denoted by $\beta\prime$, obtained from Equation (8) by dividing the complexity of the task by the computing capacity. In the case of partial reuse of computations in the cache, the cost of transferring partial input and output, as obtained from Equation (6), is also taken into account. As a result, the computation cost Equation (11) is obtained.

$$\varphi(t) = {w_{tsk}}/{\Delta_{tsk}} \quad (7)$$

$$\beta\prime = \frac{\sum_{i=1}^{f\prime} w_{\text{sub\_tsk(i)}}}{C_{CN}\Delta_{tsk}} \quad (8)$$

$$\beta = \frac{\sum_{i=1}^{x} w_{\text{sub\_tsk(i)}}}{C_{CN}\Delta_{tsk}} \quad (9)$$

$$R_\tau^{CN} = P_S \times l + (1 - P_S)(l + \beta\prime) \quad (10)$$

$$R_\tau^{CS} = P_S \times l + (1 - P_S)(l + \alpha_\tau^{part\prime} + \beta\prime) \quad (11)$$

### 4.1.5 Queuing Cost

A queue is created for each router due to congestion. It is assumed that the arrival rates of computation requests from users are modeled as bursty and with different rates. This model is presented in the form of a Markov-modulated Poisson process (MMPP). The MMPP is a multiplicative stochastic process that is an extended Poisson process by varies randomly over time. Therefore, the MMPP can describe how the input rate changes with the environment, which is more realistic. The MMPP is defined by two parameters: $Q$ and $\lambda$. $Q$ is the generator, and $\lambda$ is the Poisson entry rate for m-state, which is defined as (12) [23].

$$Q = \begin{bmatrix} -\sigma 1 & \sigma 12 & \cdots & \sigma 1m \\ \sigma 21 & -\sigma 2 & & \sigma 2m \\ \vdots & & \ddots & \vdots \\ \sigma m1 & \sigma m2 & \cdots & -\sigma m \end{bmatrix}$$

$$\lambda = (\lambda_1, \lambda_2, \cdots, \lambda_m)^T \quad (12)$$
$$\Lambda = diag(\lambda^T)$$

The service time in CN follows the exponential distribution. In our system, the rate of user requests is considered with three states: low, medium, and high. Therefore, the queue model is MMPP(3)/M/C/C.

In the article [24], it is stated that this model is NP-hard, so an approximation is used to solve it, where the M/M/C queue model serves as an appropriate approximation for the MMPP(3)/M/C queue model. To evaluate the two models, MMPP(3)/M/C and M/M/C, approximately, the queue length and waiting time were first calculated, considering the same number of servers, service rates, and average arrival rates for both models. The results indicate that the queue length and waiting time are approximately the same in both models. Furthermore, using an F-test, the results indicate that there is no significant difference in waiting time between these two models.

Considering this approximation for the model M/M/C, the equilibrium probabilities related to the number of customers in the system are examined, represented by $p_n$. The equilibrium probabilities $p_n$ indicate the probability of having n customers in the system in steady state. The flow balance equations for these queues are derived from the flow diagram, where equilibrium holds for the transitions between n−1 and n, resulting in the recursive Equation (13). The probability that a task (customer) has to wait is denoted by $w$, which is also referred to as the delay probability. The Equations for $p_0$ and the delay or waiting time $w$ for the M/M/C/C model are derived from relations (14) and (15) in [25].

$$p_n = \frac{(c\,\rho)^n}{n!} p_0 \quad n = 0, \cdots, c$$
$$p_{c+n} = \rho^n\, p_c = \rho^n \frac{(c\,\rho)^c}{c!} p_0 \quad n = 0,1,2,\cdots \quad (13)$$
$$p_0 = \left(\sum_{n=0}^{c-1} \frac{(c\,\rho)^n}{n!} + \frac{(c\,\rho)^c}{c!} \frac{1}{(1-\rho)}\right)^{-1} \quad (14)$$
$$w = p_c + p_{c+1} + p_{c+2} + \cdots$$
$$= \frac{(c\,\rho)^c}{c!}\left((1-\rho)\sum_{n=0}^{c-1} \frac{(c\,\rho)^n}{n!} + \frac{(c\,\rho)^c}{c!}\right)^{-1} \quad (15)$$

$\rho$ is the usage rate of CN, which is obtained for the M/M/C/C model from Equation (16) where $\lambda$ is the overall rate of requests entering the system, and $\mu$ represents the service rate. The service rate is the ratio of bandwidth to the size of data input and output [22]. If the bandwidth exceeds the arrival rate, the system can process data without delay; otherwise, it may encounter congestion and delays.

For our model, the Equations for the service rate and CN usage rate are (17) and (18) respectively. By inserting (17) and (18) into Equation (15), Equation (19) is obtained.

$$\rho = \frac{\lambda}{c\mu} \quad (16)$$
$$\mu = 1/\alpha_\tau^{CN} \quad (17)$$
$$\rho = \frac{\lambda\, \alpha_\tau^{CN}}{c} \quad (18)$$
$$w = \frac{\left(\frac{\lambda}{\mu}\right)^c}{c!}\left((1-\frac{\lambda}{c\mu})\sum_{n=0}^{c-1}\frac{\left(\frac{\lambda}{c\mu}\right)^n}{n!} + \frac{\left(\frac{\lambda}{\mu}\right)^c}{c!}\right)^{-1} \quad (19)$$

Users send input data for the service. If the size of the input data is large, it cannot be sent in only one packet with a length $P$ (bits) but must be transmitted in several packets. Consequently, the number of packets ($\sigma$) sent can be calculated using Equation (20), which is obtained by dividing the total input data by the length of a packet:

$$\sigma = \frac{input_{tsk}}{P} \quad (20)$$

Since there can be as many as $\sigma$ packets, there is a waiting queue at each node. Therefore, the waiting time for all incoming data can be obtained from Equation (21). The same applies to the output data.

$$w_{total} = w \times \sigma \quad (21)$$

### 4.1.6 Total Cost

According to the articles [6-7,9], there are two types of ICN request-response models for computation, which are based on pull and push. In both models, the user first sends a request that includes the size of the input data to the CN. Once the CN receives all the input data, it then executes the service. At last, the result reaches the user from the CN. In other words, the total round trip time of the computation is the sum of the time taken for input transfers, output transfers, execution time, and the waiting time of the entire queue. Several modes have been created to determine the execution time. These modes include execution from the beginning in CN, reuse of computations fully or partially in CN, or reuse of computations fully or partially in CS. The total cost, as given in Equation (22), is the sum of these modes. $P_{CN}$ represents the probability that the task is executed in CN, and $j$ is a binary parameter. when $j = 0$, it means it is executed from the beginning.

When a request arrives at the router, the router searches for the content in the CS. If the request is not found, it is sent to the next router. This process can lead to receiving packets with the same content from different locations, each having different round-trip times, which impacts delivery performance. To estimate the number of requests that have passed through each intermediate node, the impact of caching on delivery is modeled. For this analysis, the cache miss probabilities at the intermediate nodes along the request paths need to be provided. To model the delivery performance, the average virtual round-trip time of class $k$, denoted as $VRTT_K$, is defined [26]. $VRTT_K$ is the average time that elapses between sending a request packet and receiving the result packet in a steady state. $C_{TOTAL}$ represents the sum of round-trip delays.

$$C_{TOTAL} = P_{CN}\,(\alpha_\tau^{CN} + (1-j)\,\epsilon_{CN} + j\,R_\tau^{CN}) + \left((1-P_{CN})(P_S\,\alpha_\tau^{CS}) + R_\tau^{CS}\right) + w \times \sigma \quad (22)$$

In Equation (23), $1 - p_k(i)$ and $p_k(i)$ represent the collision rate and cache miss rate of class $k$ at node $i$, respectively. The requests that are answered at router $i + 1$ indicate that no collisions occurred at the previous routers $i$. The miss probability at the $i + 1$th router is equal to the product of the error probabilities at all prior routers.

$$VRTT_{total} = \sum_{i=1}^{n} c_{total} \ (1 - p_k(i)) \prod_{j=1}^{i-1} p_k(i) \quad (23)$$

Accordingly, the miss probability at router $i + 1$ is calculated iteratively using Equation (24). $p_k(1)$ is the error rate at the first router for the LRU replacement method, as shown in Equation (25).

The cache miss rate depends on the popularity of computation requests, represented by $q_k$, which indicates the popularity of content in class $k$. The cache size of the router (the number of contents that can be stored in the cache) is denoted by $x$. The cache miss rate can also be defined as the probability of the number of distinct requests in class $k$ that are not available in the cache within a given time frame. $g$ is considered an index for the number of distinct requests related to the arrival rate of requests, content class $k$, average content size, popularity distribution, and the number of computations in each class, which is calculated in Equation (26). These formulas are stated as lemmas, and their proofs are presented in the paper [24].

$$\log p_k(i) = \log p_k(1) \prod_{l=1}^{i-1} p_k(l) \quad (24)$$

$$p_k(1) = e^{-\frac{\lambda}{m} q_k g x^{\alpha}} \quad (25)$$

$$g = 1 \Big/ \lambda \, c \, \sigma^{\alpha} \, m^{\alpha-1} \Gamma(1 - \frac{1}{\alpha})^{\alpha} \quad (26)$$

## 5. Evaluation

In this section, the evaluation is presented in two parts. The first part includes investigating the construction time of the HNSW algorithm parameters, checking the accuracy of the modeling, and evaluating the search time and completion time of the method. The second part is for comparison with previous works. This comparison involves comparing the delay of getting the library descriptor and the lookup time delay of the methods. To evaluate our method, we utilized a real-world image dataset from the MNIST [27] dataset as the task input data. The size of the dataset is 40,000 contents ($M$) divided into 10 classes ($k$), with each class containing 4000 unique content types. The length of the packets is 10 KB ($P$). Each experiment was performed 10 times, and the average results were reported.

### 5.1 An Overview of the Architecture

The proposed method is implemented using the NDN architecture in the ndnSIM simulator [28]. For the experiments, we used two topologies, as illustrated in Fig. 4 and Fig. 5.

Fig. 4 consists of five nodes: a requester, three NDN routers, and an edge server (CNs) with 10ms delays set for each path. The caching capacity for the CN1 server and routers is 12000 and 100, respectively. In Fig. 5, there are five nodes, including two routers and two CNs. The cache capacity for CN1, CN2, and the routers is 10000, 12000, and 100, respectively. Additionally, the distances between nodes are different. The total round-trip delay of each route is considered to be twice the propagation delay. These topologies are designed to cover a variety of scenarios and allow for comparison and evaluation of performance under different conditions.

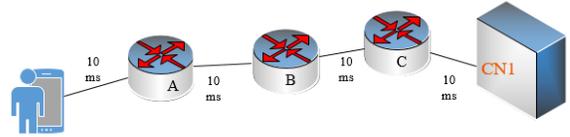

**Fig. 4.** Topology 1 for modeling and simulator accuracy verification

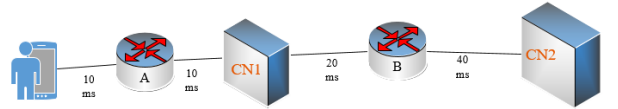

**Fig. 5.** Topology 2 for experiments

### 5.2 Experiments

Experiments have been conducted to study the effects of M(Mmax) and efConstruction parameters on the construction time of the HNSW algorithm, as shown in Fig. 5. First, we set EF to 10 and vary M to 3, 5, 7, and 9.

The Average M(Mmax) construction time for HNSW is illustrated in Table 2. Increasing M(Mmax)

results in storing more neighbors in each node, which can lead to higher search accuracy but comes with the cost of increased storage and memory consumption.

Another experiment has also been conducted with M(Mmax) equal to 5, where we vary EF to 10, 20, and 30. The average efConstruction time for HNSW is illustrated in Table 3. Higher EF values can lead to more accurate search results but result in longer construction times. Finally, a trade-off has been made between cost and accuracy, setting EF=10 and M(Mmax) =5 for the remaining experiments.

**Table 2:** Effect of the M(Mmax) construction time on M variation

| M(Mmax) value | M=9 | M=7 | M=5 | M=3 |
|---|---|---|---|---|
| The average M(Mmax) construction time(s) | 0.285 | 0.279 | 0.2725 | 0.2623 |

**Table 3:** Effect of the EF construction time on EF variation

| EF value | EF=30 | EF=20 | EF=10 |
|---|---|---|---|
| The average EF construction time (s) | 0.276 | 0.269 | 0.267 |

An experiment with the topology of Fig. 5 was conducted by varying the parameter $\alpha$ from 0.3 to 0.9 to analyze the average search time and the average completion time using the HNSW algorithm. The results are shown in Tables 4 and 5, respectively.

**Table 4:** The effect of $\alpha$ value on average search time

| Value $\alpha$ | Average Search Time(S) |
|---|---|
| $\alpha = 0.3$ | 0.176 |
| $\alpha = 0.5$ | 0.175 |
| $\alpha = 0.7$ | 0.171 |
| $\alpha = 0.9$ | 0.168 |

As the value of $\alpha$ increases, the probability of popular data also increases, resulting in reduced time to find the desired data and decreased computation time due to computation reuse. Specifically, as $\alpha$ increases, the likelihood of encountering popular data during the search process rises. Consequently, the number of hops required to reach the result in the HNSW algorithm decreases. This reduction in hops means that the overall search time decreases. As a result, the time to find the desired data decreases, ultimately leading to a reduced computation completion time.

**Table 5:** The effect of $\alpha$ value on average completion time

| Value $\alpha$ | Average Completion (S) Time |
|---|---|
| $\alpha = 0.3$ | 7.86 |
| $\alpha = 0.5$ | 7.69 |
| $\alpha = 0.7$ | 7.5 |
| $\alpha = 0.9$ | 7.44 |

To evaluate the model's accuracy, two cascade topologies are used, as illustrated in Fig. 4 and Fig. 5. Scenario 1 corresponds to the topology in Fig. 4, while scenario 2 corresponds to the topology in Fig. 5. The accuracy of the proposed model is assessed through numerical calculations and simulations. To verify the accuracy of the modeling, different values of the parameters are estimated for both the numerical calculations presented in Equation (22) and the simulations. In the figures (Fig. 6 to Fig. 13), the line labeled "proposed model" shows the modeling results derived from the formulas. The total round-trip delay of each path is considered to be twice the propagation delay.

To evaluate the modeling, the Zipf distribution function for content requests is considered, with the Equation $q_k = \frac{c}{k^\alpha}$.

Initially, the evaluation was performed under the same conditions with a Poisson distribution featuring an arrival rate of 40 requests per second and a bandwidth of 10Mbps, while the parameter $\alpha$ varies from 0.3 to 0.9 in both scenarios. The results of simulation and numerical calculations for different values of parameter $\alpha$, obtained by changing the number of tasks, have been conducted for both topologies.

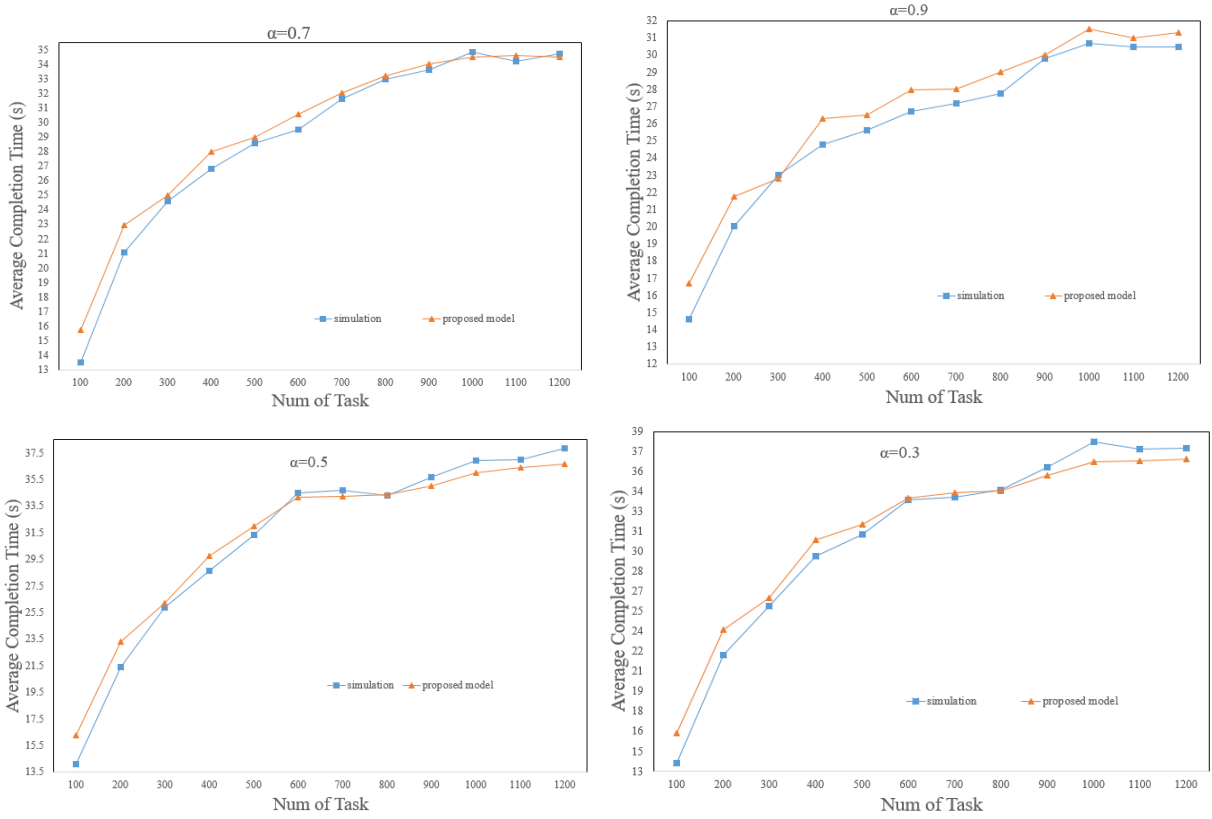

**Fig. 6.** Average completion time for different values of $\alpha$ for scenario 1

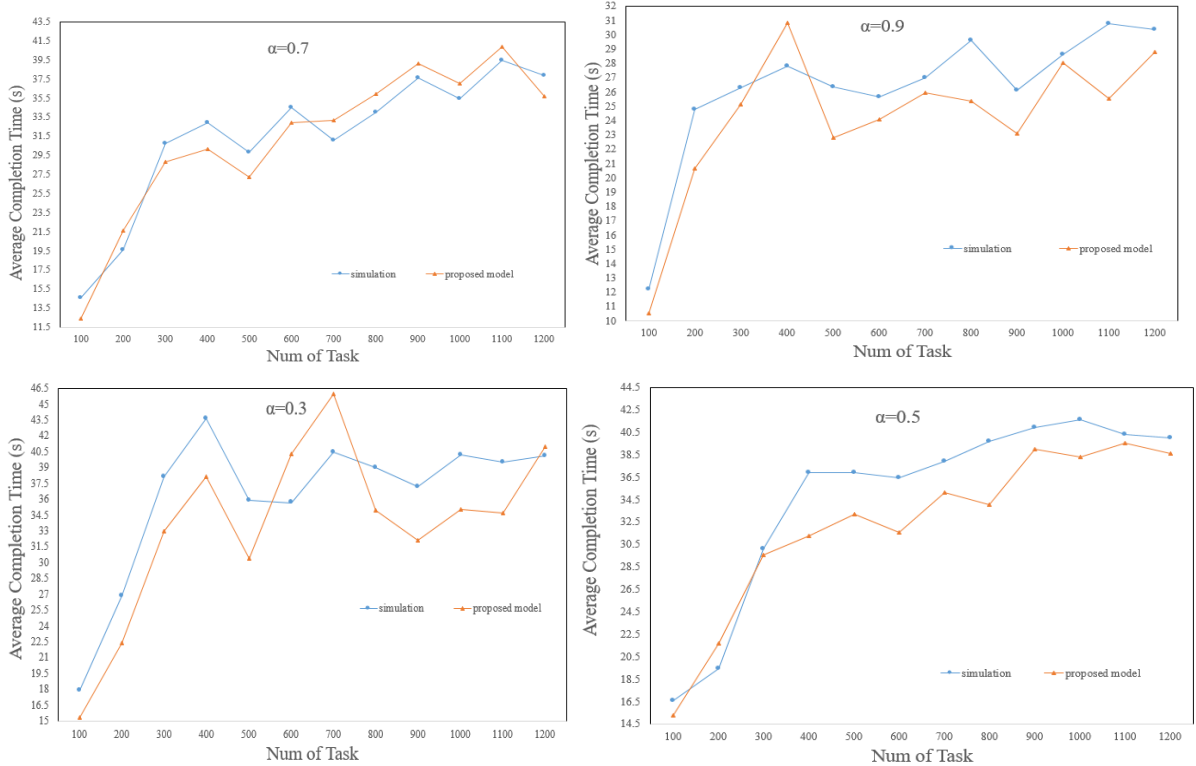

**Fig. 7.** Average completion time for different values of $\alpha$ for scenario 2

Fig. 6 illustrates the average completion time for scenario 1 and Fig. 7 illustrates the average completion time for scenario 2.

In scenario 2, due to the presence of an edge node on the critical path and the additional load on it, with the number of requests increasing, a longer completion time is observed than in scenario 1. The graphs in scenario 2 have fluctuations and peaks and troughs. This could be due to the computational complexity in scenario 2 with the presence of an edge node on the critical path.

As the value of $\alpha$ increases, the probability of using more popular data increases, leading to increased computation reuse and reduced completion time in both scenarios.

The results show that the highest percentage of error in the proposed model occurs in scenario 1 with the following values: $\alpha$=0.3 (16.6%), $\alpha$=0.5 (15.6%), $\alpha$=0.7 (16.4%) and $\alpha$=0.9, (14.2%). In scenario 2, the highest percentage of error is also observed with the following values: $\alpha$=0.3 (16.5%), $\alpha$=0.5 (15.2%), $\alpha$=0.7 (15.02%), and $\alpha$=0.9 (16.9%).

The next experiment, performed under the same conditions with a Poisson distribution featuring an arrival rate of 10 requests per second and a bandwidth of 10Mbps at $\alpha$=0.5, involved doubling the memory (packets) in both scenarios. Fig. 8 illustrates the average completion time for scenario 1, while Fig. 9 illustrates it for scenario 2.

In both scenarios, increasing the memory (packets) from 12,000 to 24,000 reduces the completion time. The effect of increasing memory (packets) is greater in scenario 2 than in scenario 1; this is because scenario 2 has more CNs, and the increase in memory (packets) on the critical path leads to a reduction in completion time.

The results indicate that the highest percentage of error in the proposed model for scenario 1, with memory (packets) of 12,000 and 24,000, is 1.3% and 9.5%, respectively. In scenario 2, the highest percentage of error in the proposed model for memory (packets) of 12,000, 10,000, and 24,000, 20,000 is 12.3% and 10.5%, respectively.

The next experiment, performed under the same conditions with a bandwidth of 10Mbps at $\alpha$=0.5, involved changing the Poisson distribution with arrival rates of 40, 20, and 10 requests per second in both scenarios.

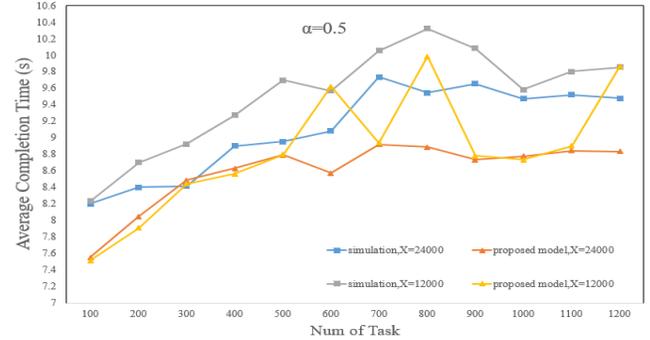

**Fig. 8.** The effect of increased (Packet) memory on completion time in scenario 1

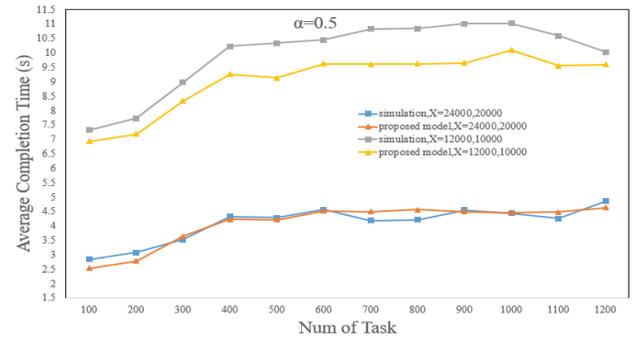

**Fig. 9.** The effect of increased (Packet) memory on completion time in scenario 2

Fig. 10 illustrates the average completion time for scenario 1, while Fig. 11 illustrates it for scenario 2.

As the arrival rate ($\lambda$) in the Poisson distribution increases in both scenarios, the completion time also increases because the number of incoming requests to the CN per second rises. In scenario 2, the completion time increases more than in scenario 1 with the rise in arrival rate ($\lambda$). This is due to the bottleneck present in this scenario, which requires simultaneous computing and data transfer. As the arrival rate increases, resource contention occurs, leading to increased computing time.

The results indicate that the highest percentage of error in the proposed model for scenario 1, using a Poisson distribution at arrival rates of 40, 20, and 10, is 15.6%, 9%, and 1.3%, respectively, while in scenario 1, it is 15.2%, 14.6%, and 12.4%, respectively.

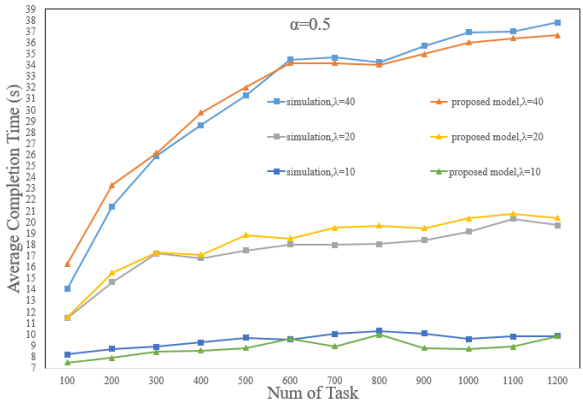

**Fig. 10.** The effect of changing the Poisson distribution on completion time in scenario 1

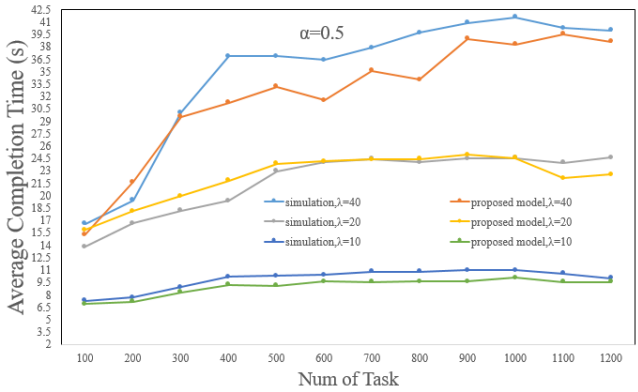

**Fig. 11.** The effect of changing the Poisson distribution on completion time in scenario 2

The next experiment, performed under the same conditions with a Poisson distribution featuring an arrival rate of 20 requests per second and $\alpha=0.5$, involved changing the bandwidth for both scenarios.

Fig. 12 illustrates the average completion time for scenario 1, while Fig. 13 illustrates it for scenario 2.

As bandwidth increases in both scenarios, the completion time decreases because the data transfer speed increases and the likelihood of packet loss or delay decreases. In scenario 5, the impact of increased bandwidth is more significant; this is because the bottleneck at the edge node on the critical path is reduced, allowing data to be transferred more quickly and with less delay.

Additionally, in scenario 5, the edge node must simultaneously handle computing and data transfer. This leads to resource contention and increased computing time, ultimately resulting in a longer completion time compared to scenario 1.

The results indicate that the highest percentage of error in the proposed model for scenario 1 at bandwidths of 1 megabit per second and 10Mbps is 11.9% and 9%, respectively, while in scenario 2, it is 12.4% and 14.6%, respectively.

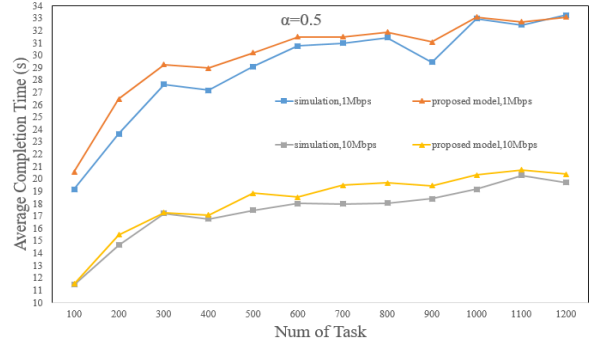

**Fig. 12.** The effect of changing bandwidth on completion time in scenario 1

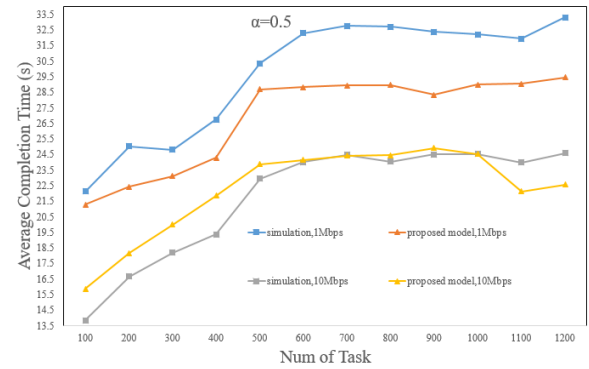

**Fig. 13.** The effect of changing bandwidth on completion time in scenario 2

### 5.3 Analysis of Completion Time and Overhead

According to the reference topology [4], which is composed of a range of 20 to 40 nodes interconnected through links with an individual delay of 5ms and with the bandwidth of 10Gbps, 10 nodes are randomly selected as ENs.

In reference [4], a 4-byte index is used for each LSH table, with one LSH table for the MNIST dataset. Additionally, there are a series of delays, including the following:

1. The delay for an NDN sender to process and send a task via FIB and rFIB is 71-101 μs and 106-74 μs, respectively.

2. The hash time delay for 1 LSH table is 0.4 milliseconds.

3. The LSH lookup time delay for 1 LSH table, as

reported in Table 6.

4. The time delay for Tensor Flow-based machine learning models for image processing is between 70 and 100 milliseconds.

While the overhead of our method includes obtaining the descriptor and implementing the HNSW algorithm, which incurs an overhead of 1ms for descriptor generation. Another overhead is the HNSW lookup, which is listed in Table 6.

Fig. 14 presents a comparative evaluation of the completion time for tasks performed in our proposed method and methods presented in [2] and [4] (Reservoir and ICedge) across all MNIST datasets. Our approach, based on the descriptor and HNSW, recognizes image similarity, significantly increasing computation reuse and resulting in reduced completion time.

**Table 6:** Search time comparison

| Number of Images (x1000) | Search time | |
|---|---|---|
| | HNSW (ms) | 1 LSH Table (ms) |
| 20 | 0.017 | 0.09 |
| 40 | 0.023 | 0.10 |
| 60 | 0.026 | 0.11 |
| 80 | 0.029 | 0.13 |
| 100 | 0.03 | 0.22 |

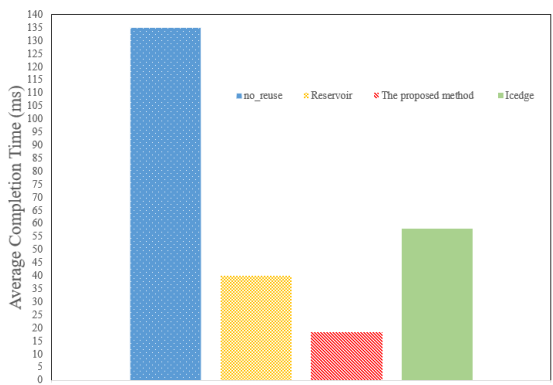

**Fig. 14.** Comparison of the proposed method with previous works

In terms of completion time, our proposed method outperforms both ICedge and Reservoir algorithms. Specifically, the completion time is 68% lower compared to ICedge and 53% lower compared to Reservoir. When compared to the no-reuse mode, our method achieves an impressive reduction of approximately 86% in completion time.

Furthermore, the completion time of our proposed method is substantially lower when compared to the no-reuse mode, with a reduction factor of 7.32x. In comparison, the reduction factors for Reservoir and ICedge are 3.375x and 2.32x, respectively, highlighting the superior performance and efficiency of our approach.

## 6. Conclusion and Future Works

In this study, the concept of the "Similarity Index" is utilized for the input parameters of images. The OpenCV library is leveraged to generate descriptors, and the HNSW algorithm is employed for efficient searching of similar descriptors. This approach allows for the utilization of similar images that produce the same computation results. The completion time of this approach is 7.32 times shorter compared to the no-reuse mode. Furthermore, a modeling approach for computing request transfer in ICN-based edge computing is introduced. The simulation results show that the highest error rate of the proposed model is approximately 16%.

In future research endeavors, there is potential to explore the development of improved descriptors that further enhance the Similarity Index. Additionally, it would be beneficial to create a Similarity Index algorithm specifically tailored for input data types such as videos, expanding the applicability and effectiveness of the proposed approach. Moreover, adding forwarding algorithms that facilitate computation reuse is recommended. Furthermore, it is important to consider more complex and multi-path topologies for modeling. In this context, addressing mechanisms for congestion control in routers is also essential.

**Declaration of generative AI and AI-assisted technologies in the writing process**

During the preparation of this work, the authors used SciSpace for paraphrasing in the introduction section and related work, and GPT-3.5 Turbo for grammar checking.